\batchmode
\documentclass[english,prd,showkeys,preprint,showpacs,elide]{revtex4-1}
\usepackage[T1]{fontenc}
\usepackage[latin9]{inputenc}
\usepackage{verbatim}
\usepackage{amsmath}
\usepackage{graphicx}
\usepackage{amssymb}

\makeatletter

\providecommand{\tabularnewline}{\\}

\@ifundefined{textcolor}{}
{%
 \definecolor{BLACK}{gray}{0}
 \definecolor{WHITE}{gray}{1}
 \definecolor{RED}{rgb}{1,0,0}
 \definecolor{GREEN}{rgb}{0,1,0}
 \definecolor{BLUE}{rgb}{0,0,1}
 \definecolor{CYAN}{cmyk}{1,0,0,0}
 \definecolor{MAGENTA}{cmyk}{0,1,0,0}
 \definecolor{YELLOW}{cmyk}{0,0,1,0}
 }

\newcommand{\dk}{\mathrm{d}\hspace{-.5pt}k}

\makeatother

\usepackage{babel}

\begin{document}

\title{Cold quark stars from hot lattice QCD}

\author{R.~Schulze, B.~Kämpfer}

\affiliation{Forschungszentrum Dresden-Rossendorf, PF 510119, 01314 Dresden, Germany\\
 and\\
 TU Dresden, Institut für Theoretische Physik, 01062 Dresden,
Germany}

\begin{abstract}
Based on a quasiparticle model for $\beta$ stable and electrically
neutral deconfined matter we address the mass-radius relation of pure
quark stars. The model is adjusted to recent hot lattice QCD results
for 2 + 1 flavors with almost physical quark masses. We find rather
small radii and masses of equilibrium configurations composed of cold
deconfined matter, well distinguished from neutron or hybrid stars.
\end{abstract}


\maketitle

\section{Introduction}

After growing evidence for the quark-gluon substructure of hadrons
the question has been asked \cite{Ito70,Bay76,Kei76,Fre78,Fec78}
whether massive neutron stars may have a core composed of quarks \citep{Gle97,*Gle00,Web99,*Web05}%
{}. These so-called hybrid stars may be part of the neutron star branch
or constitute a separate stable branch of high-density objects --
the so-called third family \citep{Ger68,Kam81a,*Kam81b,*Kam83,*Kam85}%
{} or twin stars \cite{ScB02}. Also pure quark stars populating another
separate branch of stable, spherically symmetric cold objects have
been discussed \cite{Fec78,Ana79,Pes00}. All these possibilities
depend sensitively on the equation of state at high density and the
details of the deconfinement transition at low temperature. While
at high temperature and zero net baryon density a proper numerical
evaluation of the equation state based on first-principles -- QCD
-- is accomplished, the knowledge of the equation of state at high
baryon density and low temperature is fairly poor. In the asymptotic
region, safe statements on the matter states can be made \citep{Ris01,Raj01,*SW99}%
{}, but the extrapolation to the interesting region of energy densities
around $10^{15}$ g/cm$^{3}$ is hampered by serious uncertainties
as one expects significant non-perturbative effects.

A possibility to approach the theoretical analysis of quark stars
is to employ certain models adjusted to high-temperature lattice QCD
results at zero or small net baryon density. Of course, the applicability
of such models at low temperatures and high densities is not guaranteed.
Quarks stars or neutron stars with quark cores are expected to have
similar mass-radius relations as ordinary neutron stars. This makes
difficult an experimental verification via these observables. The
modified cooling behavior of quark matter is considered as a possible
tool to find appropriate observational hints \cite{Bla00}.

Here we are interested in the mass-radius relation of pure quark stars
which are cold and spherically symmetric. We rely on a quasiparticle
model (cf.~\citep{Pes94,*Pes96,SW01,*TSW04,Gar09}%
{} for such models) which we adjust to recent realistic lattice QCD
results. Our quasiparticle model \cite{Pes94,BKS07a,Sch08} allows
for a suitable parametrization of lattice QCD data at zero and non-zero
chemical potential. Its structure can be derived from a two-loop $\Phi$
functional \cite{Pes98a,BIR01,Sch09}. To accommodate further non-perturbative
effects the running coupling $g_{s}$ is replaced by an effective
coupling $G$. In the simplest version the imaginary parts of the
self-energies are neglected and the dispersion relation is approximated
by utilizing the asymptotic self-energy. The model has been shown
to describe successfully various lattice QCD data at zero chemical
potential, at non-zero (including also purely imaginary) chemical
potential of bulk thermodynamical quantities up to off-diagonal susceptibilities
\cite{Blu04b,Blu08a,Blu08b}.

New high-temperature lattice QCD data for almost physical quark masses
\cite{Che07,Baz09} are now at our disposal for zero chemical potential.
We adjust our model at this data and extrapolate the equation of state
to zero temperature. The emerging equation of state is then used to
consider cold pure quark stars. Analog studies have been performed
in, e.g., \citep{Pes01b,*Pes03,Iva05,Fra01,*Fra02,AS02,SLS99}%
{}, however without such intimate contact to advanced lattice QCD results.

Our paper is organized as follows. In section \ref{sec:QPM} we formulate
our model for zero temperature. The comparison with hot lattice QCD
results is performed in section \ref{sec:QPMT}. The parameters are
used in section \ref{sec:quarkEOS} to gain the cold equation of state.
The emerging mass-radius relations of cold equilibrium configurations
are discussed in section \ref{sec:TOV}. The summary can be found
in section \ref{sec:Summary}. The Appendix lists expressions used
for transferring the hot lattice QCD data to finite baryon densities.

\section{Quasiparticle model at $\boldsymbol{T=0}$}

\label{sec:QPM}

For the employed quasiparticle model the pressure $p=\sum_{i=u,d,s}p_{i}$
and quark densities $n_{i}$ at temperature $T=0$ are given by 

\begin{eqnarray}
p_{i}(\mu_{i}) & = & \frac{d_{i}}{6\pi^{2}}\int_{0}^{\sqrt{\mu_{i}^{2}-m_{i}^{2}}}dk\frac{k^{4}}{\sqrt{k^{2}+m_{i}^{2}}}-B_{i}(\mu_{i}),\label{eq:pressure}\\
B_{i}(\mu_{i}) & = & B_{i}(\mu_{0})+\frac{d_{i}}{4\pi^{2}}\int_{\mu_{0}}^{\mu_{i}}d\bar{\mu}\,\frac{\partial m_{i}^{2}(\bar{\mu})}{\partial\bar{\mu}}\int_{0}^{\sqrt{\bar{\mu}^{2}-m_{i}^{2}}}dk\frac{k^{2}}{\sqrt{k^{2}+m_{i}^{2}}},\label{eq:bagpressure}\\
n_{i}(\mu) & = & \frac{d_{i}}{6\pi^{2}}(\mu_{i}^{2}-m_{i}^{2})^{3/2}\label{eq:density}\end{eqnarray}
with the index $i$ denoting the quarks $u$, $d$ and $s$ with degeneracies
$d_{u,d,s}=2N_{c}=6$. The choice of $\mu_{0}$ and the corresponding
integration constant $B(\mu_{0})$ is described below. The energy
density follows from $e=\sum_{i=u,d,s}(p_{i}+\mu_{i}n_{i})$. The
asymptotic quark masses, which enter the employed dispersion relations
$\omega_{i}^{2}=k^{2}+m_{i}^{2}$ as approximation of the self-energies,
are \begin{eqnarray}
m_{i}^{2} & = & m_{i,0}^{2}+2m_{i,0}M_{i}+2M_{i}^{2},\label{quark_mass}\\
M_{i}^{2} & = & \frac{C_{\text{f}}}{8}\left(T^{2}+\frac{\mu_{i}^{2}}{\pi^{2}}\right)G^{2}\end{eqnarray}
with $C_{\text{f}}=(N_{c}^{2}-1)/(2N_{c})$, where the rest masses
$m_{i,0}$ may be included accordingly. We employ $m_{u,0}=m_{d,0}=m_{s,0}/10$
with $m_{s,0}$ = 105 MeV as in \cite{Baz09,Che07}. For later use
also the temperature dependence is displayed here and we note already
the gluon $g$ asymptotic mass at finite temperature \begin{equation}
m_{g}^{2}=\left(\negmedspace\frac{C_{\text{b}}}{6}T^{2}+\frac{N_{c}}{12\pi^{2}}\sum_{i=u,d,s}\mu_{i}^{2}\right)G^{2},\label{gluon_mass}\end{equation}
where $C_{\text{b}}=N_{c}+\frac{3}{2}$ and the degeneracy factor
$d_{g}=N_{c}^{2}-1$.

The five relations for charge neutrality \begin{equation}
\frac{2}{3}n_{u}(\mu_{u})-\frac{1}{3}n_{d}(\mu_{d})-\frac{1}{3}n_{s}(\mu_{s})-n_{e}(\mu_{e})-n_{\mu}(\mu_{\mu})=0,\label{eq:sidecondition1}\end{equation}
 for $\beta$ equilibrium \begin{equation}
\mu_{d}=\mu_{u}+\mu_{e}\end{equation}
 (e.g., from $n\leftrightarrow p^{+}+e^{-}+\bar{\nu}_{e}$), for equilibrium
due to strangeness changing weak decays \begin{equation}
\mu_{s}=\mu_{d}\end{equation}
 (e.g., from $\Lambda\leftrightarrow p^{+}+\pi^{-}$), for $\mu$
decay \begin{equation}
\mu_{\mu}=\mu_{e}\end{equation}
 (e.g., from $\mu^{-}\leftrightarrow e^{-}+\bar{\nu}_{e}+\nu_{\mu}$),
and for total baryon density \begin{equation}
n(\mu)=\frac{1}{3}(n_{u}(\mu_{u})+n_{d}(\mu_{d})+n_{s}(\mu_{s}))\label{eq:sidecontition5}\end{equation}
map the various chemical potentials on one independent baryon chemical
potential $\mu$ via $\mu_{u,d,s,e,\mu}(\mu)$ as required as consistency
condition of the utilized quasiparticle model. We assume that the
neutrinos $\nu_{e,\mu}$ left the star matter and, therefore, do not
participate in the chemical equilibrium reactions. The pressure and
density expressions for the electron $e$ and muon $\mu$ components
are as Eqs.~(\ref{eq:pressure}) (without the functions $B_{i}$)
and (\ref{eq:density}).

The effective coupling $G^{2}$ follows from the flow equation \cite{Pes00,BKS07a}
\begin{equation}
a_{T}\frac{\partial G^{2}(\mu,T)}{\partial T}+a_{\mu}\frac{\partial G^{2}(\mu,T)}{\partial\mu}=b,\label{eq:flow}\end{equation}
where the coefficients $a_{T}$, $a_{\mu}$ and $b$ (cf.~Appendix
\ref{sec:Coefficients-Appendix}) depend again on the effective coupling
$G^{2}$ as well as both temperature and chemical potential. It is
integrated using the method of characteristics. Along each characteristic
line, the input information $G^{2}(T)$, extracted from lattice QCD
data in the next section, is transported from the temperature axis
to the chemical potential axis thus providing $G^{2}(\mu)$. Along
one arbitrary characteristic, emerging at $T=T_{0}$ and meeting the
$\mu$ axis at $\mu_{0}$, the meanfield contribution $B$ is integrated
(as outlined in Appendix \ref{sec:Coefficients-Appendix} too), yielding
the necessary integration constant $B(\mu_{0})$. With the effective
coupling $G^{2}(\mu)$ and the $B(\mu_{0})$, all thermodynamic quantities
along $T=0$ are then determined.

\section{Equation of state from lattice QCD data at $\boldsymbol{\mu=0}$}

\label{sec:QPMT}

In \cite{Baz09} (\cite{Che07}) the interaction measure $\Delta(T)/T^{4}\equiv(e-3p)/T^{4}$
has been presented for almost physical quark masses for the two light
quarks and a strange quark in the temperature range $T=140$ - 475
MeV (140 - 825 MeV) at $\mu_{i}=0$. We rely here on the p4 and asqtad
data for $N_{\tau}=8$ \cite{Baz09} and 6 \cite{Che07} and assume
that further cut-off effects are negligible, i.e.~we compare our
continuum model with the finite-size results \cite{Baz09,Che07}.
It seems most appropriate to adjust our parameters directly at the
interaction measure, being the primary information from lattice QCD,
which reads in the quasiparticle model \cite{BKS07a} for $\mu=0$
\begin{eqnarray}
\frac{e-3p}{T^{4}} & = & \frac{1}{T^{4}}\sum_{i=g,u,d,s}\left(4B_{i}+\frac{d_{i}m_{i}^{2}}{\pi^{2}}\int_{0}^{\infty}dk\, k^{2}\frac{(k^{2}+m_{i}^{2})^{-1/2}}{e^{\sqrt{k^{2}+m_{i}^{2}}/T}+S_{i}}\right),\label{eq:em3pQPM}\\
B_{i}(T) & = & B_{i,0}-\frac{d_{i}}{2\pi^{2}}\int_{T_{0}}^{T}dT'\,\frac{\partial m_{i}^{2}(T')}{\partial T'}\int_{0}^{\infty}dk\, k^{2}\frac{(k^{2}+m_{i}(T')^{2})^{-1/2}}{e^{\sqrt{k^{2}+m_{i}(T')^{2}}/T'}+S_{i}},\label{eq:Bi}\end{eqnarray}
where $S_{g}=-1$ and $S_{u,d,s}=1$ and $m_{i}$ from Eqs.~(\ref{quark_mass})
and (\ref{gluon_mass}) with explicit and implicit $T$ dependencies.
The latter one is in $G^{2}$ for which we choose \begin{equation}
G^{2}(T)=\frac{16\pi^{2}}{\beta_{0}\ln\xi^{2}}\label{eq:effcoupling}\end{equation}
with $\xi\equiv\frac{T-T_{s}}{\lambda}$ as well as $\beta_{0}=11-\frac{2}{3}N_{f}$
for $N_{f}=2+1$ flavors as a convenient parametrization of the effective
coupling which resembles a regularized 1-loop running coupling. 

Using general thermodynamic relations one can calculate the pressure
via $p(T)/T^{4}=p(T_{0})/T_{0}^{4}+\int_{T_{0}}^{T}dT'\,\Delta(T')\, T'^{-5}$,
where $p(T_{0})$ is an integration constant. The chosen value of
$T_{0}$ should be at the lower limit of our model for deconfinement,
i.e.\ $T_{0}\approx190$ MeV according to \cite{Baz09}. The scaled
entropy density is accordingly $s/T^{3}=4p/T^{4}+\Delta/T^{4}$; unfortunately,
it depends also on the pressure normalization via $4p(T_{0})/T_{0}^{4}$. 

Our quasiparticle model is based primarily on the entropy density
$s$, i.e.~pressure and interaction measure are analytic integrals
of the entropy including an integration constant $B_{0}=\sum B_{i,0}$.
Thus fitting the interaction measure means not only determining the
parameters $T_{s}$ and $\lambda$ of the effective coupling but also
the pressure integration constant $B_{0}$. Thus, our pressure as
well as entropy and energy density follow directly from $e-3p$ without
another integration constant. (This is due to additional knowledge
of explicit expressions for all thermodynamic quantities, as opposed
to general thermodynamic relations, where an additional constant $p(T_{0})$
is required to arrive from the interaction measure at the pressure.
Within the quasiparticle model, $p(T_{0})$ is known from the parameters
$T_{s}$, $\lambda$ and $B_{0}$ via the expressions of $s$ and
$\Delta$ in $p(T_{0})=(Ts(T_{0};T_{s},\lambda)-\Delta(T_{0};T_{s},\lambda,B_{0}))/4$.)

\begin{table}
\begin{tabular}{lccccc}
action  & $N_{\tau}$  & $T_{s}$ {[}MeV{]} & $\lambda$ {[}MeV{]} & $-B_{0}$ & $\chi^{2}/$dof \tabularnewline
\hline 
p4  & 6 & ~167~ & ~20~ & ~-(78 MeV)$^{4}$~ & ~0.821\tabularnewline
p4  & 8 & ~146~ & ~31~ & ~(163 MeV)$^{4}$~ & ~0.679 \tabularnewline
asqtad  & 6 & ~131~ & ~45~ & ~(90 MeV)$^{4}$~ & ~0.981 {*})\tabularnewline
asqtad  & 8 & ~107~ & ~61~ & ~(166 MeV)$^{4}$~ & ~0.654\tabularnewline
\end{tabular}

\caption{Parameters of fits to the lattice data \cite{Baz09,Che07} for $T>190$
MeV. $B_{0}=\sum_{i=g,u,d,s}B_{i,0}$.\protect \\
{*}) Without the data point at 213 MeV which would drive the fit
to fail in the high-temperature region.\label{tab:qpmparam}}

\end{table}

\begin{figure}
\includegraphics[scale=0.96]{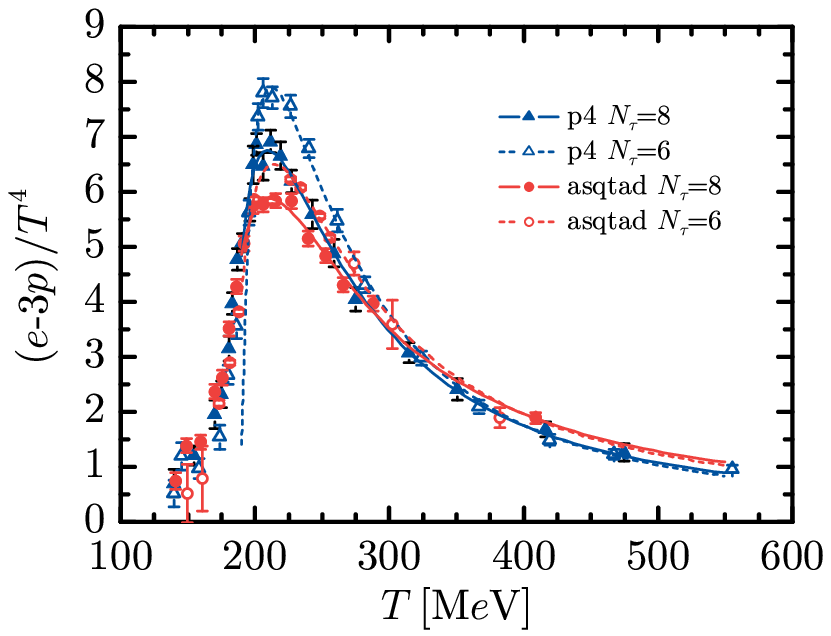}\includegraphics[scale=0.96]{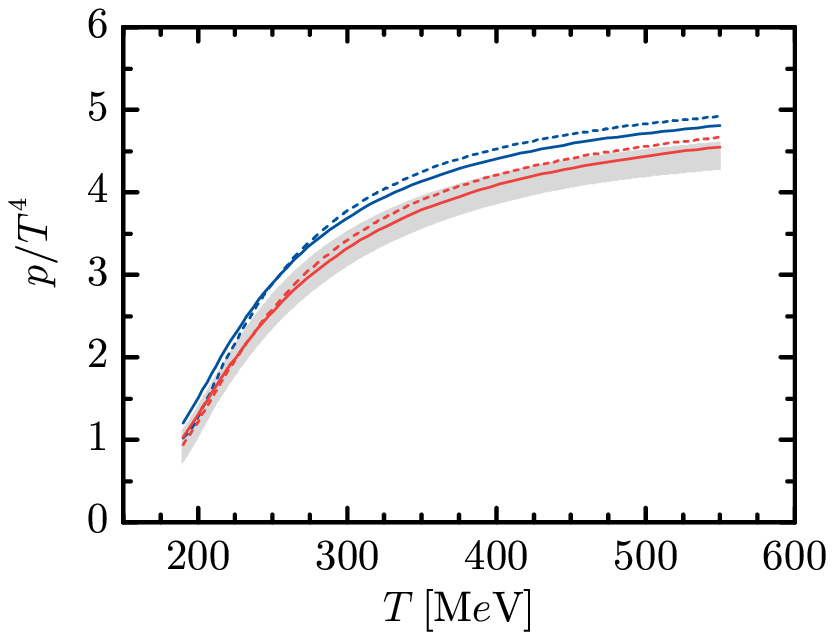}

\caption{(color online) Left panel: Fits of the quasiparticle model interaction
measure Eq.~(\ref{eq:em3pQPM}) to the lattice QCD data (symbols)
for lattice actions p4 (blue) and asqtad (red) and lattice spacings
$N_{\tau}$ from \cite{Baz09,Che07}. Solid (dashed) curves are for
$N_{\tau}=8$ (6). Right panel: Scaled pressure $p/T^{4}$ of the
quasiparticle model adjusted in the left panel compared to the pressure
estimate from \cite{Baz09} (grey area).\label{fig:pressure}\label{fig:em3p}}

\end{figure}

The $\chi^{2}$ minimization of the difference of the data in \cite{Baz09,Che07}
to the interaction measure Eq.~(\ref{eq:em3pQPM}) directly yields
the values of $T_{s}$, $\lambda$ and $B_{0}$ as listed in Tab.~\ref{tab:qpmparam}.
With this parametrization we get the interaction measure as exhibited
in the left panel of Fig.~\ref{fig:em3p}. The maximum of $\Delta$
arises from a turning point of the scaled pressure as a function of
$\log\, T$. Within our quasiparticle model, the location of the maximum
is governed by the values of $T_{s}$ and $\lambda$, where the latter
one also affects the peak width. The peak height of $\Delta/T^{4}$
is essentially determined by $B_{0}$. The fits are in a narrow corridor
for $T$ > 300 MeV yielding some confidence in the equation of state
there.

In the right panel of Fig.~\ref{fig:pressure} we compare the pressure
of our model with the pressure estimate deduced in \cite{Baz09} from
the interaction measure. Despite of the variation of the peak heights
in $\Delta/T^{4}$, the resulting pressures in our model are in a
reasonably narrow corridor: Our fit to the asqtad $N_{\tau}=8$ data
is in the middle of the pressure range determined in \cite{Baz09}
by different interpolations on the data for $\Delta/T^{4}$ and assumptions
on $p_{0}$. The p4 $N_{\tau}=6$ peak in $\Delta/T^{4}$ is higher,
and, consequently, our pressure is also somewhat higher, governed
by the positive $B_{0}$. A fit to the upper and lower limits of the
pressure band from \cite{Baz09} would result in $(T_{s},\lambda,-B_{0})$
= (109 MeV, 53 MeV,(185 MeV)$^{4}$) and (36 MeV, 107 MeV, (197 MeV)$^{4}$),
respectively.

\section{The quark equation of state at zero temperature\label{sec:quarkEOS}}

\begin{figure}
\includegraphics[scale=0.96]{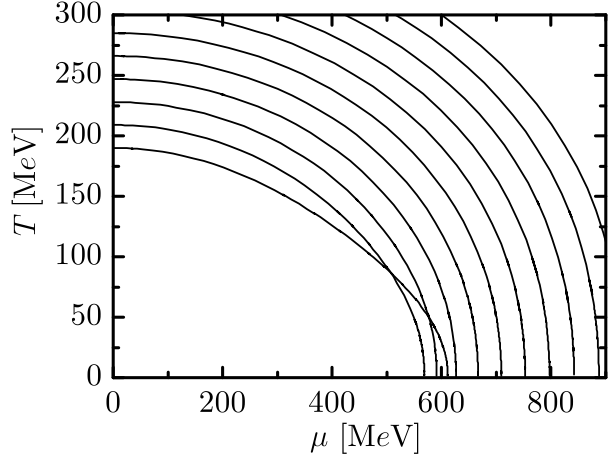}

\caption{Characteristics of the flow equation (\ref{eq:flow}) with side conditions
(\ref{eq:sidecondition1}-\ref{eq:sidecontition5}) imposed for the
p4 action and $N_{\tau}=8$.\label{fig:chars}}

\end{figure}

Utilizing the values of Tab.~\ref{tab:qpmparam} the flow equation
(\ref{eq:flow}) is solved using the mentioned method of characteristics
to obtain $G^{2}(\mu)$ and $B(\mu)$. In doing so, the side conditions
(\ref{eq:sidecondition1}-\ref{eq:sidecontition5}) are invoked so
that along each characteristic curve the requirements of $\beta$
stability and electric charge neutrality are fulfilled. The characteristics
for the p4 action and $N_{\tau}=8$ are exhibited in Fig.~\ref{fig:chars}.
As already noted in \cite{Pes00,BKS07a}, the characteristics emerging
from the very vicinity of $T_{0}$ have the tendency to cross each
other at low temperatures. (An extended version of the model in \cite{Sch08,Sch08b}
cures this insanity.) The pressure becomes negative at $\mu<$ 550
MeV. Clearly, we consider only the region of positive pressure where
the characteristics behave regularly.

It happens that the effective coupling $G^{2}$ can also be parametrized
at vanishing temperature using Eq.~(\ref{eq:effcoupling}) but with
$\xi\equiv\frac{\mu-\mu_{s}}{\lambda_{\mu}}$. The parameters are
listed in Tab.~\ref{tab:G2paramAtTeq0}. The interaction measure
$\Delta(\mu)/\mu^{4}$ displays a peak, as $\Delta(T)/T^{4}$ does.

\begin{figure}
\includegraphics[scale=0.95]{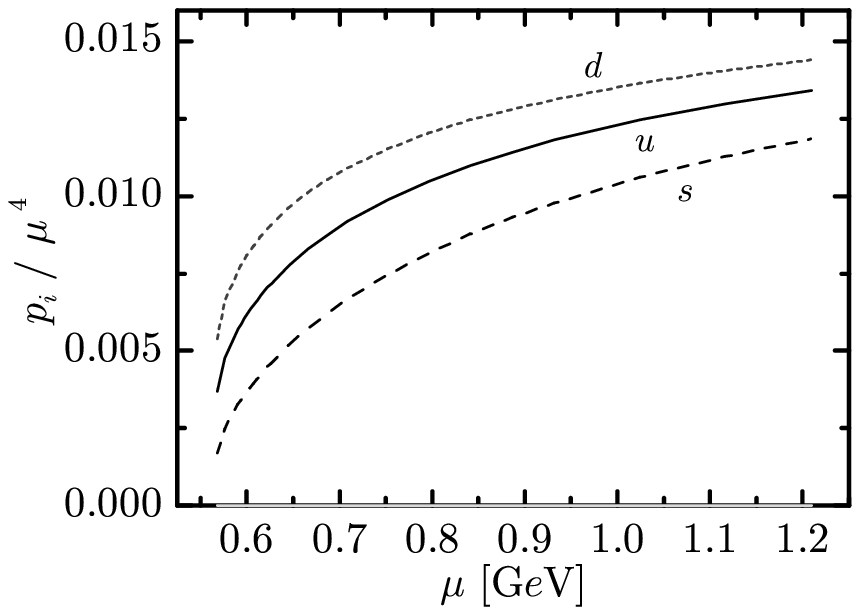}~~~~\includegraphics[scale=0.95]{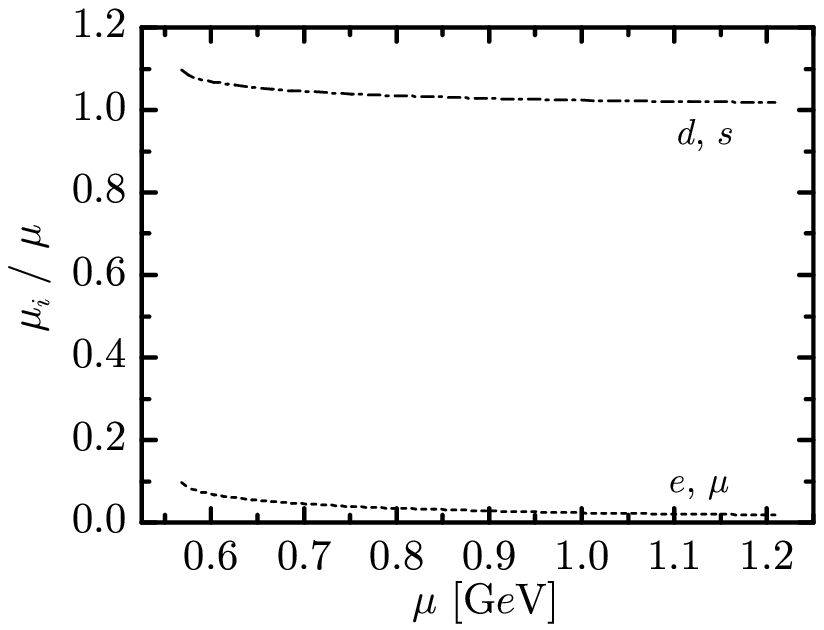}

\caption{Left panel: The scaled pressure contributions $p_{i}/\mu^{4}$ for
the p4 action and $N_{\tau}=8$ as functions of $\mu$. The leptonic
contributions are on the lower $\mu$ axis in the given $p_{i}/\mu^{4}$
scale. Right panel: The individual chemical potentials $\mu_{i}$
as functions of $\mu=\mu_{u}$.\label{fig:pressure and mue at Teq0}}

\end{figure}

\begin{figure}
\includegraphics[scale=0.93]{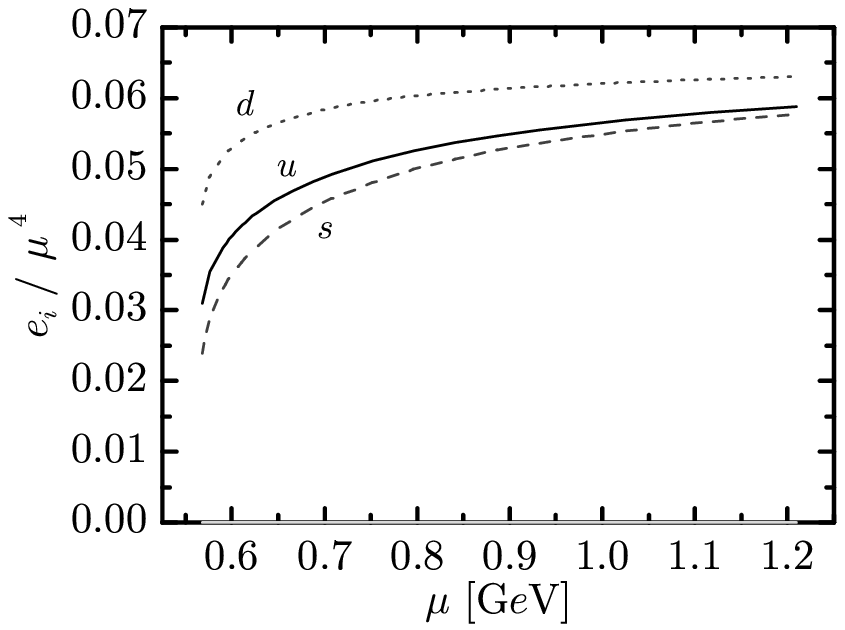}~~~~\includegraphics[scale=0.93]{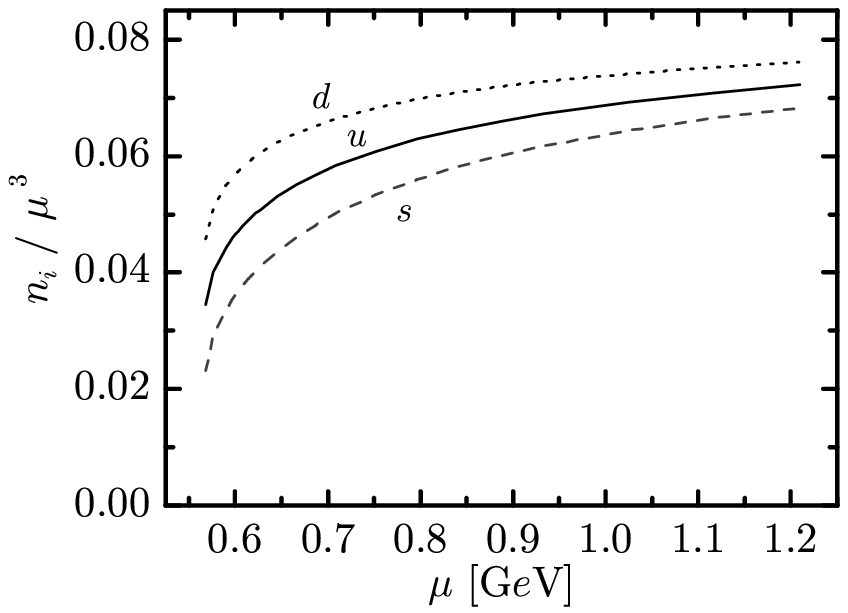}

\caption{The scaled energy density contributions $e_{i}/\mu^{4}$ (left panel)
and the scaled net quark density $n_{i}/\mu^{3}$ (right panel) for
the p4 action and $N_{\tau}=8$ as functions of $\mu$. For the used
scale, the leptonic contributions are not visible. \label{fig:energy and particle dens at Teq0}}

\end{figure}

The pressure contributions according to Eq.~(\ref{eq:pressure})
with the such obtained effective coupling $G^{2}(\mu)$ are exhibited
in the left panel of Fig.~\ref{fig:pressure and mue at Teq0} as
a function of the chemical potential $\mu=\mu_{u}$. The differences
of up, down and strange quark contributions are determined by differences
in the respective chemical potentials as shown in the right panel
of Fig.~\ref{fig:pressure and mue at Teq0}. Due to equilibrium with
respect to strangeness changing weak decays, $\mu_{d}=\mu_{s}$ holds
which deviates slightly from $\mu_{u}$. In line with \cite{SW99}
the lepton contributions are tiny, as evidenced in Fig.~\ref{fig:pressure and mue at Teq0},
too. The pressure difference of down and strange quarks is due to
the considerably larger rest mass of the latter ones. For the sake
of completeness we also show the individual contributions to energy
density (left panel in Fig.~\ref{fig:energy and particle dens at Teq0})
and the individual particle densities (right panel in Fig.~\ref{fig:energy and particle dens at Teq0}).
Similar to the pressure, the lepton contributions are not visible
on the used scales.

\begin{table}
\begin{tabular}{lccc}
action  & $N_{\tau}$  & $\mu_{s}$ {[}MeV{]} & $\lambda_{\mu}$ {[}MeV{]}\tabularnewline
\hline 
p4  & 6 & ~211~ & ~159\tabularnewline
p4  & 8 & ~134~ & ~215\tabularnewline
asqtad  & 6 & ~68~ & ~285\tabularnewline
asqtad  & 8 & ~-72~ & ~380\tabularnewline
\end{tabular}

\caption{Parameters of Eq.~(\ref{eq:effcoupling}) with $\xi\equiv\frac{\mu-\mu_{s}}{\lambda_{\mu}}$
following from the solution of the flow equation Eq.~(\ref{eq:flow}).
The fits apply in the range $\mu$ = 0.6...1.2 GeV.\label{tab:G2paramAtTeq0}}

\end{table}

\begin{figure}[h]
\includegraphics[scale=0.96]{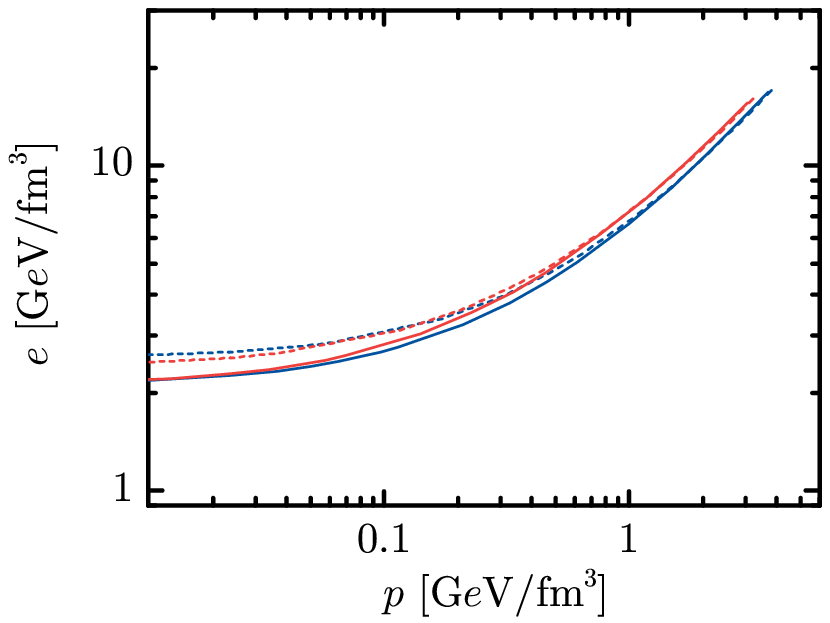}~~~~\includegraphics[scale=0.96]{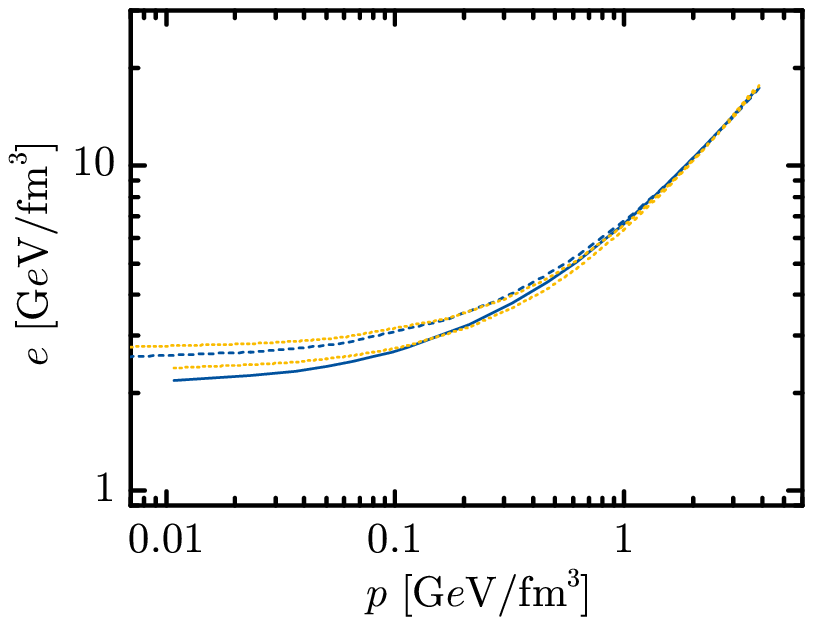}

\caption{(color online) Left panel: The energy density as a function of the
pressure at $T=0$ (solid curves) for all four considered lattice
results from \cite{Baz09,Che07}. Blue (red) lines represent results
deduced from p4 (asqtad) lattice data while solid (dashed) curves
denote $N_{\tau}=8$ (6). Right panel: Two of the linear fits $e=v_{s}^{-2}p+e_{0}$
(dotted curves) from Tab.~\ref{tab:EOSfit} compared to the corresponding
equations of state (color code as in left panel).}

\label{fig:EoS} 
\end{figure}

The resulting equation of state at $T=0$ in the form $e(p)$, needed
for the integration of the TOV equations below, is exhibited in Fig.~\ref{fig:EoS}
(left panel for a comparison of the four equations of state) together
with two fits by $e=v_{s}^{-2}p+e_{0}$ for the equations of state
adjusted to $N_{\tau}=6$ and 8 for the p4 action (right panel). Parameters
for all actions and temporal lattice extends considered here are listed
in Tab.~\ref{tab:EOSfit}. Both, the vacuum energy density $e_{0}=e(p=0)$
and the velocity of sound parameter $v_{s}^{2}=\partial p/\partial e$
are in narrow intervals for the fours sets of lattice QCD input data:
While the vacuum energy density varies within (366 MeV)$^{4}$ - (381
MeV)$^{4}$, $v_{s}^{-2}$ is within 3.8 - 4.5. As the interaction
measure $\Delta(\mu)/\mu^{4}$ becomes small at large values of $\mu$,
our resulting equations of state $e(p)$ have the tendency to merge.
(Some differences are caused by the different fit values of $B_{0}$.)
At small pressure the deviation of our four equations of state are
about 10\%.

The same is true if considering again the upper and lower limits of
the pressure band from \cite{Baz09} instead of the interaction measure.
We find values ($v_{s}^{-2},e_{0}^{1/4})$ = (3.92, 370 MeV) and (4.24,
383 MeV) in the parameter area of the above fits where larger values
of the pressure at vanishing chemical potential lead to smaller values
of the vacuum energy density at vanishing temperature and the inverse
squared velocity of sound. Also, thermal effects are found to be small,
i.e.\ up to $T=50$ MeV the equation of state $e(p)$ does not change
significantly.

\begin{table}[h]
\begin{tabular}{lccc}
action  & $N_{\tau}$  & $v_{s}^{-2}$  & $e_{0}^{1/4}$ {[}MeV{]}\tabularnewline
\hline 
p4  & 6 & ~3.81~ & ~381\tabularnewline
p4  & 8 & ~4.01~ & ~366\tabularnewline
asqtad  & 6 & ~4.23~ & ~379\tabularnewline
asqtad  & 8 & ~4.47~ & ~367\tabularnewline
\end{tabular}

\caption{Parameters of linear fits $e=v_{s}^{-2}p+e_{0}$ to our equation of
state with $G^{2}(\mu)$ determined by the flow equation (\ref{eq:flow}).
The leptonic contributions are included.\label{tab:EOSfit}}

\end{table}

\begin{figure}[h]
\includegraphics[scale=0.96]{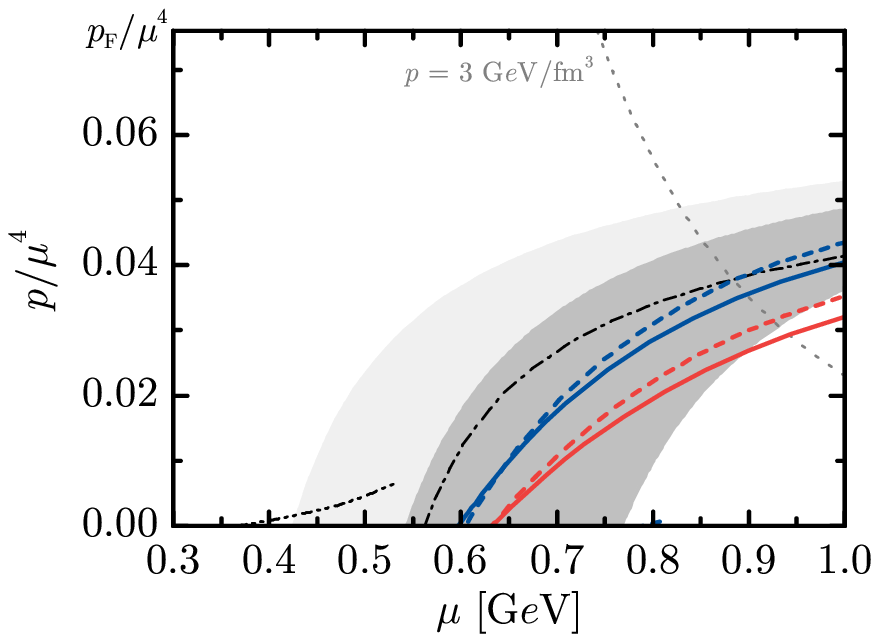}\includegraphics[scale=0.96]{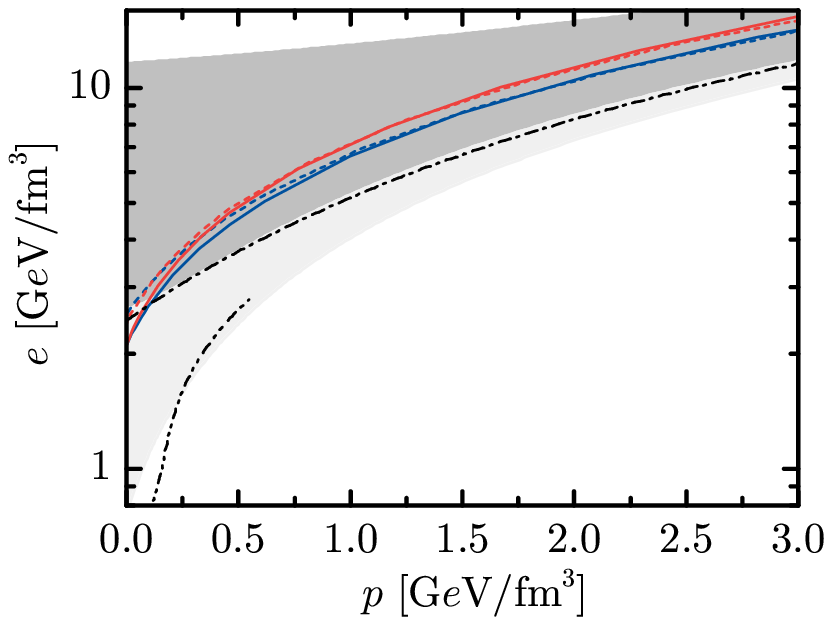}

\caption{(color online) The scaled pressure $p/\mu^{4}$ as function of $\mu$
at $T=0$ (left panel) and the equation of state $e(p)$ (right panel)
for the four considered lattice results from \cite{Baz09,Che07} in
comparison to results from \cite{AS02} (dash-dotted curve), \cite{SLS99}
(dash-double-dotted curve) and \cite{Fra01,Fra02} (grey bands limited
by $\bar{\mu}/\mu$ = 1, 1.5 and 2 from dark to light). Blue (red)
curves represent quasiparticle results adjusted to p4 (asqtad) lattice
data while solid (dashed) curves denote $N_{\tau}=8$ (6). Leptonic
contributions are included.}

\label{fig:comparisons} 
\end{figure}
Our results can be compared with perturbative calculations at vanishing
temperature in \cite{AS02,Fra01,Fra02}. In \cite{AS02} the pressure
of cold quark matter is calculated in hard-dense-loop perturbation
theory. The resulting pressure for 3 flavors with equal chemical potential
and the choice \cite{AS02} of the renormalization scale $\bar{\mu}=\mu$
is shown in the left panel of Fig.~\ref{fig:comparisons}. Also shown
are the results from the weak-coupling expansion to second order \cite{Fra01,Fra02}.
Here the value of $\bar{\mu}$ is varied from $\mu$ to 2$\mu$. For
reference also a comparison with NJL model results \cite{SLS99} is
depicted. The grey dotted curve is $p$ = 3 GeV/fm$^{3}$; the region
0 < $p$ < 3 GeV/fm$^{3}$ is relevant for quark stars, as turns out
by integrating the TOV equations (see section \ref{sec:TOV}).

The energy density $e=\mu^{2}\partial/\partial\mu(p/\mu)$ depends
on the incline of the pressure scaled with the chemical potential
rather than the absolute values of the pressure. This explains the
fact that, while the scaled quasiparticle pressure $p/\mu^{4}$ from
lattice QCD (left panel in Fig.~\ref{fig:comparisons}) shows some
spread as a function of $\mu$, the resulting equation of state $e(p)$
(right panel in Fig.~\ref{fig:comparisons}) is given as a tight
band. Inspection of $p/\mu$ as a function of $\mu$ (not displayed)
explains the broad range of values for $e(p\rightarrow0)$ in \cite{Fra01,Fra02}:
the slope of $p/\mu$ as a function of $\mu$ changes drastically
with the chosen scale $\bar{\mu}$ for small pressures. For $\bar{\mu}$
= 1.5$\mu$ the equation of state in \cite{Fra01,Fra02} in the form
$e(p)$ coincides with the results of \cite{AS02}, which in turn
falls in the same range as our set of equations of state. In fact,
$v_{s}^{-2}=3$ and $e_{0}^{1/4}$ = 365 MeV yield a good description
of the equation of state from \cite{Fra01,Fra02} for $\bar{\mu}$
= 1.2$\mu$ and \cite{AS02}. (Expanding the quasiparticle partial
pressure $p_{i}$ (\ref{eq:pressure}) including the meanfield contribution
$B_{i}$ (\ref{eq:bagpressure}) at $T=0$ in powers of the coupling
constant $G$ yields the leading terms $p_{i}(\mu_{i})=\left(1-2\alpha_{s}/\pi+\ldots\right)\mu_{i}^{4}/(4\pi^{2})+B_{i}(\mu_{0})$,
where the coefficient of the $\mathcal{O}(\alpha_{s})$ term, $\alpha_{s}=4\pi G^{2}$,
equals the strictly perturbative results in \cite{Fre77c,Fra01,Fra02};
the coefficient of the next-order term deviates from the perturbation
expansion, similar to the quasiparticle model \cite{Pes00,Pes94,Pes96}
(see also discussion in \cite{BIR01}) at non-zero temperature and
the hard-dense-loop approach in \cite{AS02}.)

Remarkable is that all the discussed equations of state have a certain
value of the chemical potential at vanishing pressure. This enables,
in principle, to construct pure quark stars with vanishing pressure
at the surface.

Fig.~\ref{fig:comparisons} clearly evidences that the previous foundation
for discussing quark stars seemed not to be on safe grounds as the
proposed model equations of state were too different unless further
constraints (as the compatibility, e.g., with a hadronic model equation
of state required in \cite{Fra02}) are imposed. Given the intimate
contact of our approach to first-principle evaluations of QCD, we
hope to have a more reliable foundation. Of course, this hope is related
to the assumption that the extrapolation to non-zero chemical potential
is sufficiently smooth. The successful comparison of our model with
Taylor expansion coefficients for the $\mu$ dependence \cite{Blu04b}
as well as the application of our model at imaginary chemical potential
\cite{Blu08a} (not only small values thereof!) give us some confidence
in our approach.

Let us finally comment on the importance of the side conditions. If
one assumes one common chemical potential for all quarks $\mu$ and
includes leptons ($\mu_{e}=\mu_{\mu})$ via a electric neutrality
condition $\mu_{e}=\mu_{e}(\mu)$ , the results of the equation of
state differ from the isospin asymmetric model with the side conditions
(\ref{eq:sidecondition1}-\ref{eq:sidecontition5}) properly invoked
on a 10\% level.

\clearpage{}

\section{Integration of the TOV equations}

\label{sec:TOV}

To estimate the properties of quark stars as spherical equilibrium
configurations of pure, strongly interacting quark matter  we employ
the TOV equations \begin{eqnarray}
\frac{dp}{dr} & = & -G_{N}\frac{([1+v_{s}^{-2}]p+e_{0})(m+4\pi r^{3}p)}{r^{2}(\,1-\frac{2m}{r}G_{N})},\\
\frac{dm}{dr} & = & 4\pi r^{2}(v_{s}^{-2}p+e_{0}),\end{eqnarray}
where the special parametrization $e=v_{s}^{-2}p+e_{0}$ of the equation
of state is supposed to hold. $G_{N}$ is the Newtonian gravitational
constant, and we employ units with $\hbar c=1$. 

\begin{figure}[h]
\includegraphics{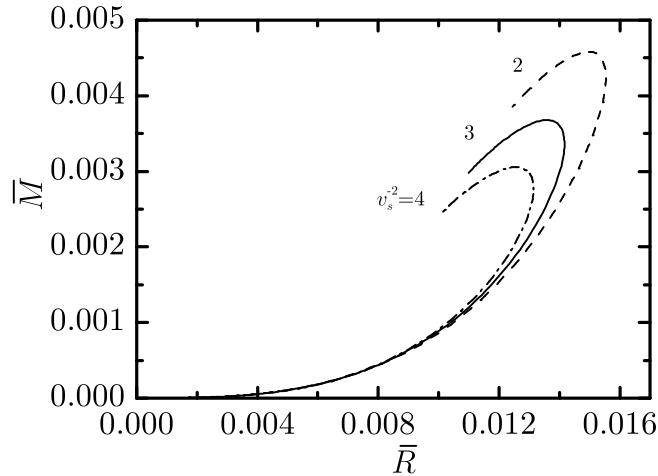}

\caption{Scaled mass $\bar{M}=\bar{m}(p=0)$ shown as a function of scaled
radius $\bar{R}=\bar{r}(p=0)$ for several values of $v_{s}^{-2}$
as solution of the scaled TOV equations (\ref{eq:TOVscaled1}) and
(\ref{eq:TOVscaled2}).}

\label{fig:scaledTOV} 
\end{figure}

We emphasize the strong dependence on the actual value of $e_{0}$
which determines the pressure gradient in the dimensionless combination
$G_{N}e_{0}^{1/2}$ (which is of the order of 10$^{-39}$ for the
case at hand), which can be seen in writing the TOV equations as\begin{eqnarray}
\frac{\partial\bar{p}}{\partial\bar{r}} & = & -\frac{\left(\left[1+v_{s}^{-2}\right]\bar{p}+1\right)\left(\bar{m}+4\pi\bar{r}^{3}\bar{p}\right)}{\bar{r}^{2}\left(1-\frac{2\bar{m}}{\bar{r}}\right)},\label{eq:TOVscaled1}\\
\frac{\partial\bar{m}}{\partial\bar{r}} & = & 4\pi\bar{r}^{2}\left(v_{s}^{-2}\bar{p}+1\right),\label{eq:TOVscaled2}\end{eqnarray}
from the scaled quantities $p=\bar{p}e_{0}$, $r=\bar{r}(G_{N}e_{0})^{-1/2}$,
$m=\bar{m}(G_{N}e_{0})^{-1/2}G_{N}^{-1}$. The scaled TOV equations
depend only on $v_{s}^{-2}$. The solutions for the relevant values
of $v_{s}^{-2}=$ 2...4 are exhibited in Fig.~\ref{fig:scaledTOV}.
With the given scaling, $m$ and $r$ shrink with increasing value
of $e_{0}^{1/2}$, while the dependence on $v_{s}^{-2}$ is moderate
within the interval covering the values of Tab.~\ref{tab:EOSfit}.
Thus the vacuum energy density $e_{0}$ is indeed the decisive quantity
determining the sizes and the masses of pure quark stars. To be specific,
for $v_{s}^{-2}=3\pm1$, the scaled maximum mass is 0.004 $\pm$ 0.001.

\begin{figure}
\includegraphics{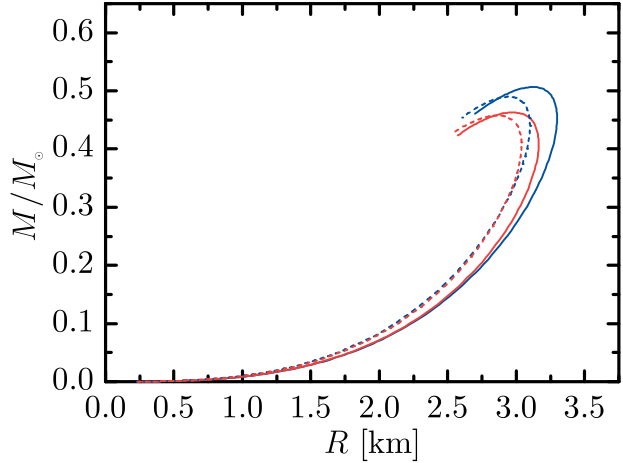}

\caption{(color online) Mass-radius relations of quark stars following directly
from lattice QCD results \cite{Baz09,Che07} within our quasiparticle
approach. Blue (red) lines represent results deduced from p4 (asqtad)
lattice data while solid (dashed) curves denote $N_{\tau}=8$ (6).}

\label{fig:qStars}
\end{figure}

To test the dependence of deviations from the approximation $e=v_{s}^{-2}p+e_{0}$
we integrate the TOV equations with our equations of state adjusted
to the lattice QCD results. The results are exhibited in Fig.~\ref{fig:qStars}.
The maximum masses are about 0.5$M_{\odot}$ with radii of about 3
km. If such objects would exist, their bulk characteristics were quite
different from canonical neutron stars with masses concentrated at
1.4 $M_{\odot}$ and radii of 15 km and larger. Therefore, the pure
quark stars from our analysis cannot serve as candidates of twin stars
discussed in \cite{ScB02}.

We stress again the important role of the value of $e_{0}=e(p=0)$.
With the above derived scaling, equations of state with significantly
smaller values of $e_{0}$ than deduced in our analysis of the lattice
QCD results combined with the employed quasiparticle model, would
allow for significantly larger masses and radii.

The present considerations will be modified when combining our equation
of state of deconfined matter with a hadronic low-density equation
of state at $p>0$. Then hybrid stars could be constructed with properties
depending to a large extent on the transition region from confined
to deconfined matter.

\section{Summary}

\label{sec:Summary}

In summary we employ a quasiparticle model, adjusted to recent realistic
lattice QCD data with almost physical quarks masses, to consider pure
quark stars. The needed equation of state can be approximated very
well by the concise form $e=v_{s}^{-2}p+e_{0}$ with values of $v_{s}^{-2}=3.8-4.5$
and $e_{0}^{1/4}=366-381$ MeV from the lattice QCD data \cite{Baz09}.
Lattice data from both the p4 and the asqtad version can be described
equally well and lead to similar spherically symmetric stars. The
maximum masses are about 0.5 ${\rm M}_{\odot}$ with radii of 3 km. 

The pure quarks star masses and radii scale with $e_{0}^{-1/2}$ which
is the decisive quantity as $e_{0}$ is the vacuum energy density
at vanishing pressure. It follows within our model directly from the
lattice QCD data at finite temperatures.

Rapidly rotating quark stars exhibit a disc like shape with sharp
edge. Their maximum masses are enlarged by 65\% and the radii by a
similar amount \cite{Ans03} at the shedding limit. 

The present approach can be extended to the full HTL quasiparticle
model \cite{Sch08,Sch09}, where effects of Landau damping and collective
modes are included

\textbf{Acknowledgments}: The work is supported by BMBF 06DR9059D.
Useful discussions with M.~Bluhm, D.~Blaschke, R.~Meinel, D.~Petrov
and A.~Teichmüller are gratefully acknowledged.

\appendix

\section{Coefficients of the flow equation\label{sec:Coefficients-Appendix}}

The coefficients in Eq.~(\ref{eq:flow}) are

\begin{eqnarray*}
a_{T} & = & -\sum_{i=u,d,s}\frac{\partial n_{i}}{\partial m_{i}^{2}}\left.\frac{\partial m_{i}^{2}}{\partial G^{2}}\right|_{T,\mu},\\
a_{\mu} & = & \sum_{i=g,u,d,s}\frac{\partial s_{i}}{\partial m_{i}^{2}}\left.\frac{\partial m_{i}^{2}}{\partial G^{2}}\right|_{T,\mu},\\
b & = & \sum_{i=g,u,d,s}\frac{\partial n_{i}}{\partial m_{i}^{2}}\left.\frac{\partial m_{i}^{2}}{\partial T}\right|_{G^{2},\mu}-\frac{\partial s_{i}}{\partial m_{i}^{2}}\left.\frac{\partial m_{i}^{2}}{\partial\mu}\right|_{G^{2},T}\end{eqnarray*}
with

\begin{eqnarray*}
\frac{\partial n_{i}}{\partial m_{i}^{2}} & = & -\frac{d_{i}}{4\pi^{2}T}\int_{0}^{\infty}\!\dk\,\frac{k^{2}}{\omega_{i}}F_{-},\\
\frac{\partial s_{i}}{\partial m_{i}^{2}} & = & -\frac{d_{i}}{4\pi^{2}T}\int_{0}^{\infty}\!\dk\, k^{2}\left\{ F_{+}-\frac{\mu_{i}}{\omega_{i}}F_{-}\right\} ,\end{eqnarray*}
where $F_{\pm}=\left[f_{+}^{2}\exp\frac{\omega_{i}-\mu_{i}}{T}\pm f_{-}^{2}\exp\frac{\omega_{i}+\mu_{i}}{T}\right]$
and $f_{\pm}$ are the statistical distribution functions for fermions
(+) and anti-fermions (-) respectively. The derivatives of the effective
gluon masses (\ref{gluon_mass}) are \begin{eqnarray*}
\left.\frac{\partial m_{g}^{2}}{\partial T}\right|_{G^{2},\mu} & = & \frac{C_{\text{b}}}{3}TG^{2}+\frac{N_{c}}{12\pi^{2}}\left(N_{l}\!+\!2N_{h}\right)\left(\mu+\mu_{e}\right)\left.\frac{\partial\mu_{e}}{\partial T}\right|_{G^{2},\mu}G^{2},\\
\left.\frac{\partial m_{g}^{2}}{\partial\mu}\right|_{G^{2},T} & = & \frac{N_{c}}{12\pi^{2}}\left(\Big(2\left(N_{l}\!+\! N_{h}\right)\mu+\left(N_{l}\!+\!2N_{h}\right)\mu_{e}\Big)G^{2}+\left(N_{l}\!+\!2N_{h}\right)\left(\mu+\mu_{e}\right)\left.\frac{\partial\mu_{e}}{\partial\mu}\right|_{G^{2},T}\right),\\
\left.\frac{\partial m_{g}^{2}}{\partial G^{2}}\right|_{T,\mu} & = & \frac{C_{\text{b}}}{6}T^{2}+\frac{N_{c}}{12\pi^{2}}\left(N_{l}\!+\!2N_{h}\right)\left(\mu+\mu_{e}\right)\left.\frac{\partial\mu_{e}}{\partial G^{2}}\right|_{T,\mu}G^{2},\end{eqnarray*}
where $N_{l}$ and $N_{h}$ are the numbers of included light (2)
and heavier (1) quark flavors. For the effective quark masses ($i=u,d,s$)
one has\begin{eqnarray*}
\left.\frac{\partial m_{i}^{2}}{\partial T}\right|_{G^{2},\mu}\!\!\! & = & \left.\frac{\partial m_{i}^{2}}{\partial T}\right|_{G^{2},\mu,\mu_{e}}\!\!\!+\left.\frac{\partial m_{i}^{2}}{\partial\mu_{e}}\right|_{G^{2},T,\mu}\left.\frac{\partial\mu_{e}}{\partial T}\right|_{G^{2},\mu},\\
\left.\frac{\partial m_{i}^{2}}{\partial\mu}\right|_{G^{2},T}\!\!\! & = & \left.\frac{\partial m_{i}^{2}}{\partial\mu}\right|_{G^{2},T,\mu_{e}}\!\!\!+\left.\frac{\partial m_{i}^{2}}{\partial\mu_{e}}\right|_{G^{2},T,\mu}\left.\frac{\partial\mu_{e}}{\partial\mu}\right|_{G^{2},T},\\
\left.\frac{\partial m_{i}^{2}}{\partial G^{2}}\right|_{T,\mu} & = & \left.\frac{\partial m_{i}^{2}}{\partial G^{2}}\right|_{T,\mu,\mu_{e}}+\left.\frac{\partial m_{i}^{2}}{\partial\mu_{e}}\right|_{G^{2},T,\mu}\!\left.\frac{\partial\mu_{e}}{\partial G^{2}}\right|_{T,\mu}\end{eqnarray*}
with $\left.\partial m_{i}^{2}/\partial T\right|_{G^{2},\mu,\mu_{e}}=VTG^{2}$,
$\left.\partial m_{i}^{2}/\partial\mu\right|_{G^{2},T,\mu_{e}}=V\mu_{i}G^{2}/\pi^{2}$,
$\left.\partial m_{i}^{2}/\partial\mu_{e}\right|_{G^{2},T,\mu}=\delta_{iu}V\mu_{i}G^{2}/\pi^{2}$
and $\left.\partial m_{i}^{2}/\partial G^{2}\right|_{T,\mu,\mu_{e}}=\left(T^{2}+\mu_{i}^{2}/\pi^{2}\right)/2$,
where $V=\left(m_{q,0}/M_{q}+2\right)C_{\text{f}}/4$. The derivatives
of the electron chemical potential therein are\begin{eqnarray*}
\left.\frac{\partial\mu_{e}}{\partial T}\right|_{G^{2},\mu} & = & -W^{-1}\sum_{j=u,d,s,e,\mu}q_{j}\left.\frac{\partial n_{j}}{\partial T}\right|_{G^{2},\mu,\mu_{e}},\\
\left.\frac{\partial\mu_{e}}{\partial\mu}\right|_{G^{2},T} & = & -W^{-1}\sum_{j=u,d,s,e,\mu}q_{j}\left.\frac{\partial n_{j}}{\partial\mu}\right|_{G^{2},T,\mu_{e}},\\
\left.\frac{\partial\mu_{e}}{\partial G^{2}}\right|_{T,\mu} & = & -W^{-1}\sum_{j=u,d,s,e,\mu}q_{j}\left.\frac{\partial n_{j}}{\partial G^{2}}\right|_{T,\mu,\mu_{e}}\end{eqnarray*}
with $W=\sum_{j=u,d,s,e,\mu}q_{j}\left.\partial n_{j}/\partial\mu_{e}\right|_{G^{2},\mu,T}$
and\begin{eqnarray*}
\left.\frac{\partial n_{i}}{\partial T}\right|_{G^{2},\mu,\mu_{e}} & = & \left.\frac{\partial n_{i}}{\partial T}\right|_{G^{2},\mu,\mu_{e},m_{i}^{2}}+\frac{\partial n_{i}}{\partial m_{i}^{2}}\left.\frac{\partial m_{i}^{2}}{\partial T}\right|_{G^{2},\mu,\mu_{e}},\\
\left.\frac{\partial n_{i}}{\partial\mu}\right|_{G^{2},T,\mu_{e}} & = & \left(\left.\frac{\partial n_{i}}{\partial\mu_{i}}\right|_{G^{2},m_{i}^{2}}+\frac{\partial n_{i}}{\partial m_{i}^{2}}\left.\frac{\partial m_{i}^{2}}{\partial\mu_{i}}\right|_{G^{2},T,\mu_{e}}\right)\left.\frac{\partial\mu_{i}}{\partial\mu}\right|_{\mu_{e}},\\
\left.\frac{\partial n_{i}}{\partial\mu_{e}}\right|_{G^{2},\mu,T} & = & \left(\left.\frac{\partial n_{i}}{\partial\mu_{i}}\right|_{G^{2},m_{i}^{2}}+\frac{\partial n_{i}}{\partial m_{i}^{2}}\left.\frac{\partial m_{i}^{2}}{\partial\mu_{i}}\right|_{G^{2},\mu}\right)\left.\frac{\partial\mu_{i}}{\partial\mu_{e}}\right|_{\mu},\\
\left.\frac{\partial n_{i}}{\partial G^{2}}\right|_{T,\mu,\mu_{e}} & = & \frac{\partial n_{i}}{\partial m_{i}^{2}}\left.\frac{\partial m_{i}^{2}}{\partial G^{2}}\right|_{T,\mu,\mu_{e}}\end{eqnarray*}
as well as\begin{eqnarray*}
\left.\frac{\partial n_{i}}{\partial T}\right|_{G^{2},\mu,\mu_{e},m_{i}^{2}}\!\!\!\!\! & = & \frac{d_{i}}{2\pi^{2}T}\int_{0}^{\infty}\!\!\!\dk\, k^{2}\left\{ \omega_{i}F_{-}-\mu_{i}F_{+}\right\} ,\\
\left.\frac{\partial n_{i}}{\partial\mu_{i}}\right|_{G^{2},m_{i}^{2}} & = & \frac{d_{i}}{2\pi^{2}T}\int_{0}^{\infty}\!\dk\, k^{2}F_{+}.\end{eqnarray*}
$q_{i}$ are the electric charges of the quark species. Note that
the side conditions Eqs.~(\ref{eq:sidecondition1}-\ref{eq:sidecontition5})
are included. These strongly modify the coefficients given in \cite{BKS07a}.

Along the characteristics, where $\mu$, $T$ and $G^{2}$ are given
as functions of the affine curve parameter $x$, the bag pressure
$B$ has to be integrated according to\[
B=B(\mu=0)-\sum_{i}\int_{0}^{x}\text{d}x\frac{\partial p_{i}}{\partial m_{i}^{2}}\left(a_{T}\left.\frac{\partial m_{i}^{2}}{\partial T}\right|_{G^{2},\mu}+a_{\mu}\left.\frac{\partial m_{i}^{2}}{\partial\mu}\right|_{G^{2},T}+b\left.\frac{\partial m_{i}^{2}}{\partial G^{2}}\right|_{T,\mu}\right)\]
with\[
\frac{\partial p_{i}}{\partial m_{i}^{2}}=-\frac{d_{i}}{4\pi^{2}}\int_{0}^{\infty}\!\dk\,\frac{k^{2}}{\omega_{i}}\left[f_{+}+f_{-}\right].\]

\end{document}